\documentclass[%
reprint,
 amsmath,amssymb,
 aps,
]{revtex4-1}

\usepackage{graphicx}% Include figure files
\usepackage{dcolumn}% Align table columns on decimal point
\usepackage{bm}% bold math
\usepackage{tikz}
\usetikzlibrary{arrows,shapes}
\usetikzlibrary{trees}
\usetikzlibrary{patterns}
\usetikzlibrary{matrix,arrows} 				% For commutative diagram
\usetikzlibrary{positioning}				% For "above of=" commands
\usetikzlibrary{calc,through}				% For coordinates
\usetikzlibrary{decorations.pathreplacing}  % For curly braces
\usepackage{pgffor}							% For repeating patterns

\usetikzlibrary{decorations.pathmorphing}	% For Feynman Diagrams
\usetikzlibrary{decorations.markings}
\tikzset{
    vector/.style={decorate, decoration={snake}, draw},
	provector/.style={decorate, decoration={snake,amplitude=2.5pt}, draw},
	antivector/.style={decorate, decoration={snake,amplitude=-2.5pt}, draw},
        smallvector/.style={decorate, decoration={snake,amplitude=1.5pt,post length=0.5mm}, draw},
    fermion/.style={draw=black, postaction={decorate},
        decoration={markings,mark=at position .55 with {\arrow[draw=black]{>}}}},
    fermionbar/.style={draw=black, postaction={decorate},
        decoration={markings,mark=at position .55 with {\arrow[draw=black]{<}}}},
    fermionnoarrow/.style={draw=black},
    gluon/.style={decorate, draw=black,
        decoration={coil,amplitude=4pt, segment length=5pt}},
    scalar/.style={dashed,draw=black, postaction={decorate},
        decoration={markings,mark=at position .55 with {\arrow[draw=black]{>}}}},
    scalarbar/.style={dashed,draw=black, postaction={decorate},
        decoration={markings,mark=at position .55 with {\arrow[draw=black]{<}}}},
    scalarnoarrow/.style={dashed,draw=black},
    electron/.style={draw=black, postaction={decorate},
        decoration={markings,mark=at position .55 with {\arrow[draw=black]{>}}}},
    bigvector/.style={decorate, decoration={snake,amplitude=4pt}, draw},
    arrow/.style={draw=black, postaction={decorate},
        decoration={markings,mark=at position 1 with {\arrow[draw=black]{>}}}},
}

\tikzstyle{block} = [draw, rectangle, 
    minimum height=3em, minimum width=6earticlem]

\def\be{\begin{equation}}
\def\ee{\end{equation}}
\def\bea{\begin{eqnarray}}
\def\eea{\end{eqnarray}}
\def\<{\langle}
\def\>{\rangle}
\def\pa{\partial}   
   
\newcommand{\eps}{\epsilon}

\newcommand{\reef}[1]{(\ref{#1})}

\begin{document}

\begin{flushright}
BOW-PH-165 ~~~
MCTP-16-28
\end{flushright}

\preprint{APS/123-QED}

\title{Soft Photon and Graviton Theorems in Effective Field Theory}

\author{Henriette Elvang}
 \email{elvang@umich.edu}
\author{Callum R.~T.~Jones}%
 \email{jonescal@umich.edu}
\affiliation{%
Michigan Center for Theoretical Physics and \\
 Randall Laboratory of Physics, Department of Physics,\\
	   University of Michigan \\ Ann Arbor, MI 48109, USA
}%

\author{Stephen G.~Naculich}
 \email{naculich@bowdoin.edu}
\affiliation{
Department of Physics \\ Bowdoin College \\
Brunswick, ME 04011, USA\\
}

\begin{abstract}
Extensions of the photon and graviton soft theorems are derived in 4d local effective field theories with massless particles of arbitrary
spin. We prove that  effective operators can result in new terms in
the soft theorems at subleading order for photons and subsubleading
order for gravitons. 
The new soft terms are unique and we provide
a complete classification of all local operators responsible for such
modifications.  We show  that 
no local operators can modify the subleading
soft graviton theorem. 
The soft limits are taken in a 
manifestly on-locus manner using a complex double deformation 
of the external momenta.  In addition to the new soft theorems, 
the resulting master formula 
yields consistency conditions such as the
conservation of electric charge, the Einstein equivalence principle,
supergravity Ward identities, and the Weinberg-Witten theorem.

% \begin{description}
% Secondary publications and information retrieval purposes.
% \item[PACS numbers]
% May be entered using the \verb+\pacs{#1}+ command.
% You may use the \texttt{description} environment to structure your abstract;
% use the optional argument of the \verb+\item+ command to give the category of each item. 
% \end{description}
\end{abstract}

\pacs{Valid PACS appear here}% PACS, the Physics and Astronomy
\maketitle

\section{\label{sec:intro}Introduction}
In this letter, we show that in a 4d local effective field theory
of only massless particles, the tree-level soft photon and graviton
theorems receive modifications at subleading and subsubleading orders,
respectively. Specifically, in effective field theory,  the soft theorems for positive-helicity  soft photons or gravitons take the form
\bea
  \label{phsoft}
   A_{n+1}^\text{ph} \!&=&\! \Big( \frac{S^{(0)}}{\eps^2}
     +\frac{S^{(1)}}{\eps} \Big) A_n 
         + \frac{\tilde{S}^{(1)}}{\eps}  \tilde{A}_n 
 + O(\eps)\,,\\
   \label{gravsoft}
 A_{n+1}^\text{grav} \!&=&\! \Big( \frac{\mathcal{S}^{(0)}}{\eps^3}
     +\frac{\mathcal{S}^{(1)}}{\eps^2}+\frac{\mathcal{S}^{(2)}}{\eps}\Big) A_n 
     + \frac{\tilde{\mathcal{S}}^{(2)}}{\eps}  \tilde{A}_n      
     + O(\eps)\,,~~~
\eea
where $S^{(i)}$ and $\mathcal{S}^{(i)}$ are the standard soft factors, well-known from the work of 
\cite{Low:1954kd,GellMann:1954kc,Low:1958sn,Weinberg:1965nx,Burnett:1967km,Gross:1968in,Jackiw:1968zza,Cachazo:2014fwa},
 and given explicitly in  \reef{photonS} and \reef{gravS} below. The new soft terms are
\be
  \label{newSs}
  \tilde{S}^{(1)}\tilde{A}_n = \sum_k g_{k}\frac{[sk] }{\<sk\>}\tilde{A}^{(k)}_n     \,,
  ~~~
  \tilde{\mathcal{S}}^{(2)}\tilde{A}_n  = \sum_k {g}_{k}\frac{[sk]^3}{\<sk\>}\tilde{A}^{(k)}_n   
\,,
\ee
where $g_k$  denotes the couplings of the associated 
effective operators. The tilde and superscript $(k)$ on the $n$-point amplitude indicate that the particle type of the 
$k$th leg of  $\tilde{A}_n$ may differ  from 
that in $A_{n+1}$. Thus, the new soft terms are different 
from the  factorized form of the traditional soft theorems. Only a small  
set of effective operators can modify the soft theorems and we provide a complete  classification. We show that no matter which operator is responsible for the modification, the kinematic soft factor is uniquely fixed to take the form \reef{newSs}. 
Our result for the photon soft theorem naturally generalizes to non-abelian gauge theory.   

Only   effective operators with 3-point interactions can affect
the single-particle soft theorems in (\ref{phsoft}) and (\ref{gravsoft}). 
If an operator has too many derivatives, its interaction is too soft to affect the soft theorems at these orders. For example,  $\text{tr}F^3$ does not 
modify the soft theorem, but the Pauli dipole operator $\overline{\chi}
\gamma^{\mu\nu} F_{\mu\nu} \chi$ does. All effective operators that can
modify the soft theorems \reef{phsoft}-\reef{gravsoft} are listed in
\reef{photonOps} and \reef{gravOps}. Note that our results imply that
the {\em soft graviton theorem is  not corrected at subleading order
$1/\eps^2$ in effective field theory}. This is important for recent
proposals  \cite{Kapec:2016jld,Cheung:2016iub}  connecting soft graviton  theorems
to asymptotic symmetries.

To investigate the soft limits, we present a novel approach based on a double complex deformation of the amplitudes. Combining a ``soft shift" with two 
BCFW shifts allows us to identify the parts of the amplitude responsible for the soft theorems as factorization poles. 
Note that we are not deriving new recursion relations and the results are independent of which lines are shifted along with the soft line.  The method allows 
us to take the soft limit in a manifestly on-locus fashion that emphasizes the  path dependence of the soft theorems at subleading order.

The  approach yields not only  the well-known soft theorems and new soft terms, it also implies non-trivial, though well known, 
consistency conditions, such as charge conservation, the equivalence principle, and the supersymmetric Ward identities which state that a spin 3/2 particle must be coupled supersymmetrically to a graviton. We also demonstrate the Weinberg-Witten theorem \cite{Weinberg:1980kq} in the form 
that no massless spin $>2$ particle can couple consistently to massless particles with spin 2 or less 
\footnote{Similar conclusions were also reached for example in \cite{McGady:2013sga} using factorization.}.

Soft theorems have  been connected to asymptotic symmetries 
\cite{Cachazo:2014fwa,Kapec:2014opa,He:2014cra,Lysov:2014csa,Kapec:2016jld,Cheung:2016iub} 
and this has led to a recent cascade of soft limit investigations,
especially at tree level. Our work was motivated by the question of possible loop corrections. This is subtle for loops of massless particles because of IR divergences, but loops of massive particles can be integrated out to leave effective operators. As we show here, certain local operators can indeed modify the soft theorems at subleading orders. It would be interesting to know if these new universal modifications are associated with asymptotic symmetries.

\section{\label{sec:cmpl}Complex deformations}

We work with spinor helicity formalism in 4d following the conventions of
\cite{Elvang:2015rqa,Elvang:2013cua}. Momenta are assumed to be 
complex so that
angle and square spinors are independent. The momentum $p_s = - |s\>[s|$ is taken soft   holomorphically: 
$|s\> \to \eps |s\>$ and $|s] \to |s]$,
with $\eps$ a small parameter.
The standard soft theorems for soft 
{\em positive-helicity} 
photons and gravitons 
then take the form \reef{phsoft}-\reef{gravsoft} (without the tilde'd modifications), where for a soft photon
\be  
  \label{photonS}
   S^{(0)} = \sum_k g_k \frac{\<xk\>}{\<xs\>\<sk\>}\,,~~~~
   S^{(1)} = \sum_k \frac{g_k}{\<sk\>} D_{sk}\,,~~~~  
\ee
and for a soft graviton
\bea
  \nonumber
   \mathcal{S}^{(0)} \!&=&\! \kappa \sum_k  \frac{[sk]\<xk\>\<yk\>}{\<sk\>\<xs\>\<ys\>}\,,~~~~
   \mathcal{S}^{(2)} = \frac{\kappa}{2} \sum_k  \frac{[sk]}{\<sk\>} D_{sk}^2\,,~~~~  \\
  \label{gravS}
   \mathcal{S}^{(1)}  \!&=&\! \frac{\kappa}{2} \sum_k  \frac{[sk]}{\<sk\>}
      \bigg( \frac{\<xk\>}{\<xs\>}+\frac{\<yk\>}{\<ys\>} \bigg) D_{sk}\,.~~~~
\eea
Here $|x\>$ and $|y\>$ are arbitrary reference spinors and $D_{sk} \equiv   |s]_a \pa_{|k]_a}$. When the amplitudes have their momentum-conserving delta functions 
stripped off, the derivatives are taken with a prescription where one uses momentum conservation to eliminate a choice of two square spinors \cite{Cachazo:2014fwa}.

In this note we use a prescription in which the soft limit is taken along a path on the algebraic locus in momentum space defined by requiring that the external momenta are on-shell and satisfy $(n+1)$-particle momentum conservation. 
Start with
$n$ 
 {\em unshifted} momenta 
$p_k = -|k\> [k|$ 
satisfying $n$-particle momentum conservation, 
$\sum_{k=1}^n p_k = 0$. 
Introduce the soft momentum $p_s = - |s\> [s|$ such that the {\em shifted}  momenta  
$\hat{p}_k = -|\hat{k}\> [\hat{k}|$,
defined as 
\be
  \label{shift}
  \begin{split}
  |\hat{s}\> &= \eps |s\> -z  |X\>\,,\\
  |\hat{i}] &= |i] - \eps \frac{\<js\>}{\<ji\>} |s] + z \frac{\<jX\>}{\<ji\>} |s] \,,\\
  |\hat{j}] &= |j] - \eps \frac{\<is\>}{\<ij\>} |s] + z \frac{\<iX\>}{\<ij\>}  |s] \,,
  \end{split}
\ee
with no other spinors shifted,
satisfy $(n\!+\!1)$-particle momentum conservation, 
$\hat{p}_s + \sum_{k=1}^n \hat{p}_k = 0$.  
 The spinor $|X\>$  
  is completely arbitrary.
  The complex deformation \reef{shift} can be viewed as the combination of a {\em soft $\eps$-shift} \cite{Cheung:2015cba} and two BCFW shifts with parameters $z_1 = \tfrac{\<jX\>}{\<ji\>} z$ and $z_2 = \tfrac{\<iX\>}{\<ij\>} z$, and 
 $z |X\> = z_1 |i\> +z_2 |j\>$.  
  The choice of the two lines $i$ and $j$ is arbitrary and does not affect the physics conclusions. 

For any momentum $k=1,\dots, n$ we have
\be \label{Pks}
  \hat{P}_{sk}^2 = (\hat{p}_k + \hat{p}_s)^2 = 
(\eps - \eps_k) P_{sk}^2 \,, 
  ~~~\eps_k = z \frac{\<Xk\>}{\<sk\>} \, .
\ee  
Evaluating $\hat{P}_{sk} $ at $ \eps = \eps_k$, 
we obtain $\hat{P}_{sk}  = - |k\> [ \hat{P}_{sk}| $ with 
\be
 \begin{split}
 &|\hat{P}_{is}] = |i]\,,~~~|\hat{P}_{js}] = |j]\,,\\
 &|\hat{P}_{sk}] = |k] + z \frac{\<Xs\>}{\<sk\>} |s] \,,~~\text{for}~~k\ne i,j\,.
 \end{split}
\ee
We are interested in poles at $\eps = 0$ in the $(n\!+\!1)$-particle amplitude. With $z=0$, there are multiple contributions to such poles, since (as is obvious from \reef{Pks}) all 2-particle channels with a soft line $s$ contribute. The role of $z\ne 0$ is to separate 
these poles to different locations in the complex $\eps$-plane and exploit that the amplitude factorizes on simple poles. Since the only possible poles in $\epsilon$ come from the 2-particle channels, we can write
\be
  \label{Anp1pre}
  \hat{A}_{n+1}(z,\eps) = \sum_{k,h_P,c} \hat{A}_3\big(\hat{s},\hat{k},-\hat{P}_{sk,c}^{h_P}\big) 
  \frac{1}{\hat{P}_{sk}^2} \hat{A}^{(k)}_n(z) + O(\eps^0)\,,
\ee
because 
when a propagator goes on shell, 
the amplitude factorizes into a product of on-shell amplitudes \footnote{In this and subsequent formulae we leave implicit any additional minus signs which may arise from reordering and crossing fermion lines during factorization or from rearrangements of spinor brackets. These signs are important for getting correct results, particularly for the relative signs in the supersymmetry Ward identities (\ref{swi}) but we will not spell them out in this Letter.}.   
The sum is over all relevant momentum channels $k$ as well as over the spectrum of particles on the internal line, as indicated with the helicity label $h_P$ and a collective index $c$ of other quantum numbers.
The superscript on the $n$-point amplitude indicates that it in general depends on the channel momentum $k$: 
$\hat{A}^{(k)}_n(z) = \hat{A}_n\big( \hat{P}_{sk,\bar{c}}^{-h_P} , \ldots\big)$.

Little-group scaling fixes the 3-particle amplitude up to a constant which we absorb in the associated coupling $g_{H_k}$, where $H_k=\{h_s, h_k, h_P; a,b,c\}$ labels helicities and possible quantum numbers:
\be
  \label{3ptA}
  \hat{A}_3\big(\hat{s},\hat{k},-\hat{P}_{sk}^{h_P}\big)
  = g_{H_k}\, 
   [\hat{s} \hat{k}]^{x_1} [\hat{k} \hat{P}_{sk}]^{x_2} [\hat{P}_{sk} \hat{s}]^{x_3}\,,
\ee
where $x_1 = h_s + h_k - h_P$, $x_2 = h_k + h_P - h_s$, and $x_3 = h_P + h_s - h_k$. 
In special 3-particle kinematics,  another option is 
that $A_3$ could depend on angle brackets only;  
however, the shifted angle brackets vanish. 
The mass dimension of the coupling is 
\be
 \label{mdimgdefa}
 [g_{H_k}]=
\mathtt{a} - 2h_s\,,
  \quad {\rm with} \quad
\mathtt{a} \equiv h_s - h_k - h_P +1 \,.
\ee
Using the kinematics above,  \reef{Anp1pre} becomes 
\be
  \label{Ann}
\hat{A}_{n+1}(z,\eps)
=\!\! \sum_{k,h_P,c} \!\!
  \frac{g_{H_k}
[sk]^{2h_s-\mathtt{a}} \<Xs\>^{1-\mathtt{a}}\,\hat{A}^{(k)}_n(z)}
  {\eps\, z^{\mathtt{a}-1} \<sk\>^{2-\mathtt{a}}\big( 1 - \frac{z}{\eps} \frac{\<Xk\>}{\<sk\>}\big)} 
  +
  O(\eps^0)\,.
\ee
This is the ``master formula''  for the following analysis. (For comments about signs, see footnote [19].) We work with the Laurent expansion \reef{Ann} for sufficiently small $z \ll \eps$ and, as we shall see, the soft theorems then follow from the $O(z^0)$ terms.

\section{Photon and graviton consistency conditions}
\label{sec:cons}
At tree level, locality requires that an amplitude can be singular only on a factorization channel. For $z=0$ and generic $\epsilon\neq 0$ there is no associated channel, so the appearance of such a pole violates locality. Therefore, if the value of $\mathtt{a}$ is greater than 1 in \reef{Ann},  the sum of residues of the apparent poles at $z=0$ must vanish.  This imposes non-trivial constraints on the amplitudes. 

Two non-trivial constraints arising from this requirement are
\be
  \label{consistency}
  \begin{split}
  h_s = 1,~\mathtt{a} =2~~&\implies~~\sum_{k} g_{H_k} = 0\,,\\
  h_s = 2,~\mathtt{a}=3~~&\implies~~\sum_{k=1}^n g_{H_k} [sk]\<sk\>= 0\,.
  \end{split}
\ee
The first condition is simply {\em charge conservation}. The second condition can  be satisfied only when the graviton couples identically to all particles; we recognize this as the {\em equivalence principle}. These results were first obtained by a different argument by Weinberg \cite{Weinberg:1964ew}.
 
We now prove that a unitary local theory can have no interactions with 
$\mathtt{a} \ge 4$.  Let the highest value of $\mathtt{a}$ in a theory be $\mathtt{a}_\text{max}\ge 4$.  The kinematic structure of the corresponding 3-particle amplitude 
$A_3\big(s_a^{h_s},1_b^{h_1},P_c^{h_P}\big)$ is uniquely determined 
by little group scaling as in (\ref{3ptA}).  
Denote the coupling by $f_{abc}$, where $a,b,c$ are collective indices for all internal quantum numbers. 
CPT invariance requires that the theory also includes the amplitude of the 
CP conjugate states; its coupling is  $f_{\overline{abc}} = f^*_{abc}$. 
Consider  the soft limit of the 4-particle amplitude  $A_4\big(s_a^{h_s},1_b^{h_1},2_{\overline{a}}^{-h_s},3_{\overline{b}}^{-h_1}\big)$, whose  $s1$-channel diagram includes the 3-particle interaction with $\mathtt{a}_\text{max}\ge 4$ and its conjugate, as well as the $s2$- and $s3$-channel diagrams, if relevant. The consistency condition arising from the absence of the pole $1/z^{\mathtt{a}_\text{max}-1}$ in \reef{Ann} implies
\be 
  \label{Bkexpr}
  \sum_{k=1}^3\langle s k\rangle^{\mathtt{a}_\text{max}-2}B_k = 0\, ,
\ee
where $B_1= \sum_{c} f_{abc} [sk]^{2h_s-\mathtt{a}_\text{max}} \hat{A}_3^{(k)}(0)$ and similarly for $B_2$ and $B_3$ 
(if present). Importantly, 
the $B_i$ are independent of  $|s\>$. 
Applying the operator $|p\rangle^{\dot{a}}\partial_{|s\rangle^{\dot{a}}}$ to \reef{Bkexpr} gives
\be
  \label{condB}
  \sum_{k=1}^3 \langle kp\rangle \langle sk\rangle^{\mathtt{a}_\text{max}-3}B_k = 0\,.
\ee
Since $|s\>$ and $|p\>$ are arbitrary, we can choose them to be $|2\>$ and $|3\>$ in which case \reef{condB} requires $B_1 = 0$. (Similarly, one can show $B_2=B_3=0$.) Since
\be
   B_{1} \propto \sum_{c}f_{abc}f_{\overline{abc}} = \sum_c |f_{abc}|^2,
\ee
it can vanish only if $f_{abc}=0$. This shows that any couplings of interactions with $\mathtt{a}\ge 4$ must vanish.

For $\mathtt{a}=3$, the above argument fails because the power of $\<sk\>$ in \reef{condB} is no longer strictly positive. Indeed, $\mathtt{a}=3$ is perfectly fine for gravitons. 
For soft photons, however, we have proven that there are no interactions with $\mathtt{a}_\text{max} > 2$.

Consider two examples of excluded interactions: \\[1mm]
$\bullet$  A 3-particle interaction with $h_s = 1 $ and $h_k = h_P = -1$ gives $\mathtt{a} =4$. 
It may appear  strange that such an interaction is excluded here, since the gluon amplitude 
$A_3(1^+, 2^-, 3^-)$ certainly exists and is non-vanishing in Yang-Mills theory. However, this 3-gluon amplitude is non-vanishing in terms of angle brackets only. To produce such an amplitude in terms of  square brackets 
only would require a non-local interaction $A^2  \tfrac{\partial}{\Box} A$ \cite{Elvang:2015rqa,Elvang:2013cua}.\\[1mm]
$\bullet$  Consider a soft  photon case of $\mathtt{a}= 3$: take $h_s =  1$, 
$h_k =-1$, $h_P=0$. This matrix element can be obtained from the operator $\Box^{-2} F_{\mu\nu} F^\nu{}_\rho \partial^\mu \partial^\rho \phi$ which clearly is not local.   

Since $h_s+h_k+h_P=0$ implies $\mathtt{a}= 2h_s+1$, we conclude from the above bounds on $\mathtt{a}$ that no 3-point interactions
involving photons, gravitinos, or gravitons are allowed if the
sum of the three helicities vanishes.

\section{\label{sec:photon}Soft photon theorems}
\noindent {\bf Standard soft photon theorem.}
Set $\mathtt{a}=2$ in  the master formula \reef{Ann} for a soft positive-helicity photon ($h_s=1$). Expanding the $n$-point amplitude and the denominator factor in small $z$, there are two contributions at order $z^0$. One goes as $1/\eps^2$ and takes the form
\be
  \label{phepsSQ}
  \hat{A}_{n+1} (z,\eps)\big|_{z^0,  {1/\eps^2} } 
=  \frac{1}{\eps^2}\sum_{k,c} g_{H_k}\frac{\<Xk\> }{\<Xs\>\<sk\>} A_n\,,
\ee
where $A_n$ is the unshifted amplitude,
which is a function of the momenta $p_k$
that satisfy $n$-particle momentum conservation. 
The result \reef{phepsSQ} is the standard leading soft factor $S^{(0)}$.

The other $O(z^0)$ contribution is order $1/\eps$:
\be
  \label{pheps}
  \hat{A}_{n+1} (z,\eps)\big|_{z^0,  
{1/\eps} 
} =  \frac{1}{\eps}\sum_{k,c} \frac{g_{H_k}}{\<Xs\>} \partial_z 
\hat{A}_n(z) \big|_{z=0}\,.
\ee
The shifted amplitude $\hat{A}_n(z)$ depends on $z$ through the momentum 
line $\hat{P}_{sk}$ as well as potentially through the 
shifted momenta $\hat{p}_i$ and $\hat{p}_j$. In the momentum channel with $\hat{P}_{sk}^2 = 0$, one uses the chain rule to find $\partial_z \hat{A}_n(z) = \tfrac{\<Xs\>}{\<sk\>} \nabla_{sk} \hat{A}_n(z)$ 
with
\be
   \nabla_{sk}
   \equiv  |s]_a  \bigg(  \pa_{|k]_a}
      + \frac{\<ki\>}{\<ij\>}  \pa_{|j]_a}
      - \frac{\<kj\>}{\<ij\>}  \pa_{|i]_a}
      \bigg)\,.
\ee 
The first term gives the familiar subleading soft factor $S^{(1)}$. 
The two other terms are consequences of our prescription for taking the soft limit. In contrast to \cite{Cachazo:2014fwa} where the soft limit is taken by defining an \textit{extrinsic} continuation of the amplitude off-locus 
(away from the support of the momentum conserving delta function), our soft limit is calculated along an on-locus path defined by the $z=0$ deformation (\ref{shift}). The corresponding soft theorems can therefore depend only on \textit{intrinsic} on-locus data. The modified differential operators can be understood as an element of tangent space of the  momentum conserving locus. Our soft limit prescription can be shown to be equivalent to that of \cite{Cachazo:2014fwa}.

\vspace{1mm}
\noindent {\bf Modification of the subleading soft photon theorem.}
The only other $1/\eps$ contributions from \reef{Ann}
for $h_s=1$ arise from interactions with $\mathtt{a}=1$. 
These give 
\be
  \label{phsubsoft}
  \hat{A}_{n+1} (z,\eps)\big|_{z^0,  
{1/\eps} 
}
  =
  \frac{1}{\eps}\sum_{k,c} g_{H_k}\frac{[sk] }{\<sk\>} \tilde{A}^{(k)}_n + O(\eps^0)\,,
\ee
which yields the new subleading soft factor $\tilde{S}^{(1)}\tilde{A}_n $
in (\ref{newSs}). By \reef{mdimgdefa},
the coupling must have mass dimension $-1$ and $h_k+h_P =1$. The new $\mathtt{a}=1$ contribution to the subleading soft theorem involves an $n$-point amplitude $\tilde{A}_n$ whose external states may differ from the $n$ hard states of $A_{n+1}$. To determine which theories can have these corrections, one simply goes through the options to find that the only possible operators are
\be
  \label{photonOps}
  \begin{split}
  &\overline{\chi} \gamma^{\mu\nu} F_{\mu\nu} \chi \,,~~~
  \phi F_{\mu\nu} F^{\mu\nu}\,,~~~
  \phi F_{\mu\nu} \tilde{F}^{\mu\nu}\,,~~~
  \\
  &\overline{\psi}_\mu F_{\nu\rho} \gamma^{\mu\nu\rho} \chi\,,~~~
  h F^2 \,,
  \end{split}
\ee
where $\chi$ is a spin 1/2 field and $\psi_\mu$ is the gravitino field. The operator $h F^2$ is shorthand for the 3-particle interaction that arises from the metric expansion of $F_{\mu\nu} F^{\mu\nu}$.

To summarize, we have shown that in effective field theory the soft  theorem for a 
positive-helicity  soft photon takes the form \reef{phsoft}, where $S^{(0)}$ and $S^{(1)}$ are as given in \reef{photonS} with $D_{sk} \to \nabla_{sk}$ on the momentum conserving locus. 
The new soft factor $\tilde{S}^{(1)}$ in \reef{newSs} 
is unique no matter which of the possible effective operators in \reef{photonOps} are responsible for the modification of the soft theorems.

%%%%
\section{\label{sec:graviton}Soft graviton theorems}

\noindent {\bf Standard soft graviton theorem.}
The familiar terms \reef{gravS} of the graviton soft theorem \reef{gravsoft} follow from the master equation \reef{Ann} by setting $h_s=2$ and $\mathtt{a}=3$. As we have already seen in \reef{consistency}, the absence of the $1/z^2$ pole in this expression implies the equivalence principle: the graviton couples uniformly to all  particles with a universal coupling $\kappa=g_{H_k}$. Using this, the  $1/z$ terms can be rewritten in terms 
of the Lorentz generators $J_{ab} = \frac{i}{2}\sum_k \big( |k]_{a} \partial_{[k|^{b}} +|k]_{b} \partial_{[k|^{a}} \big)$ and terms that vanish by momentum conservation. 
Since $J_{ab}$ 
annihilates the on-shell amplitudes, the residue of the $1/z$ pole  vanishes without imposing further constraints. 
The $O(z^0)$ terms 
give the soft theorem \reef{gravsoft} in a form with 
\bea
  \nonumber
   \mathcal{S}^{(0)} \!&=&\! \kappa \sum_k  [sk]\<sk\> 
   \frac{\<Xk\>^2
}{\<Xs\>^2\<sk\>^2}\,,~~
   \mathcal{S}^{(2)} = \frac{\kappa}{2} \sum_k  \frac{[sk]}{\<sk\>} \nabla_{sk}^2\,, \!\!\! \\
  \label{gravSnew}
   \mathcal{S}^{(1)}  \!&=&\! \kappa \sum_k  
\frac{\<Xk\>[sk]}{\<Xs\>\<sk\>}
      \nabla_{sk}\,.
\eea
Using the Schouten identity to write 
e.g.~$\frac{\<Xk\>}{\<Xs\>\<sk\>} = \frac{\<Xy\>}{\<Xs\>\<sy\>}-\frac{\<ky\>}{\<ks\>\<sy\>}$ as well as using momentum conservation and annihilation of the amplitude by  $J_{ab}$, one can show that the soft factors \reef{gravSnew} are equivalent to those in \reef{gravS} with the replacement $D_{sk} \to \nabla_{sk}$ as discussed for the photon soft theorem above. 

\vspace{1mm}
\noindent {\bf Subleading soft graviton theorem unchanged.}
The only way to get a modification to the soft graviton theorem at order $1/\eps^2$ is via interactions with $\mathtt{a}=2$. The responsible local operators would have couplings of  mass dimension $-2$ and give rise to 3-particle amplitudes $A_3\left(1^{+2},2^{h_2},3^{h_P}\right)$ with $h_2+h_P =1$. Restricting to spin$\,\le\,$2, the options are $(h_2,h_P) = (2,-1),(\tfrac{3}{2},-\tfrac{1}{2}),(1,0),(\tfrac{1}{2},\tfrac{1}{2})$. The requirement that the $1/z$ pole in \reef{Ann} vanishes implies that no such {\em local} operators exist.  For the case $(h_2,h_P) = (2,-1)$, consider the 4-graviton amplitude  
$A_4(1^{+2},2^{+2},3^{-2},4^{-2})$  at quadratic order in the non-standard effective coupling $g_c$.  Only one factorization channel contributes to $1/z$ in \reef{Ann}, namely
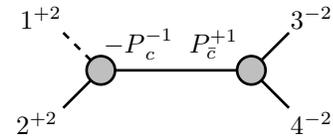
  \begin{center}
      {\begin{tikzpicture}[line width=1 pt,scale=0.5]
          \draw [scalarnoarrow] (-1,1)--(0,0);
          \draw (-1,-1)--(0,0);
          \draw (4,0)--(0,0);
          \draw (5,1)--(4,0);
          \draw (5,-1)--(4,0);          \
          \draw[black,fill=lightgray] (0,0) circle (2.5ex);
          \draw[black,fill=lightgray] (4,0) circle (2.5ex);
          \node at (-1.6,1.4) {$1^{+2}$};
          \node at (-1.7,-1.4) {$2^{+2}$};
          \node at (5.6,1.4) {$3^{-2}$};
          \node at (5.6,-1.4) {$4^{-2}$};
          \node at (1,0.7) {${-P}^{-1}_c$};
          \node at (3,0.7) {${P}^{+1}_{\bar{c}}$};
        \end{tikzpicture}} 
    \end{center}
with implicit sum over possible internal quantum numbers $c$ of the exchanged spin-1 state.
  CPT invariance requires the couplings of the two interactions to be conjugate, so the $1/z$ pole in \reef{Ann} will be proportional to $\sum_c |g_c|^2$.  Absence of this pole requires $g_c=0$.  The three other cases of $\mathtt{a}=2$ interactions can be similarly
excluded. In conclusion, in a unitary CPT-invariant theory 
there can exist no local operators that modify the subleading soft graviton theorem. 

This result may have relevance 
to recent discussions of asymptotic symmetries. In \cite{Kapec:2014opa} it was shown that the universality of the subleading soft graviton theorem (\ref{gravsoft}) is equivalent to the Ward identity of a Virasoro symmetry of the quantum gravity S-matrix. Our result implies that the subleading soft graviton theorem, and consequently the Virasoro symmetry, is unmodified at tree-level in the presence of local effective operators. In particular this includes curvature corrections.

\vspace{1mm}
\noindent {\bf Modification of the subsubleading soft graviton theorem.}
The only other $1/\eps$ contributions  from \reef{Ann} for $h_s=2$
arise from interactions with $\mathtt{a}=1$.  By \reef{mdimgdefa},
the coupling must have mass dimension $-3$ and $h_k+h_P =2$. 
The corresponding operators in effective field theory are
\be
   \label{gravOps}
   \phi R_{\mu\nu\rho\sigma}R^{\mu\nu\rho\sigma}\,,~~~
   R^{\mu\nu\rho\sigma} \overline{\psi}_\rho \gamma_{\mu\nu} \partial_\sigma \chi\,,~~~
   R_{\mu\nu\rho\sigma} F^{\mu\nu} F^{\rho\sigma}\,.
\ee 
All of these operators, up to constants, give the same correction 
$  \tilde{\mathcal{S}}^{(2)}\tilde{A}_n  $ in \reef{newSs} to the soft theorem.  
The modification due to the operator $\phi R^2$ was previously noted by \cite{Bianchi:2014gla,DiVecchia:2016amo}.

\section{\label{sec:other}
Weinberg-Witten and supergravity}

\vspace{1mm}
\noindent {\bf Weinberg-Witten.} 
The equivalence principle mandates that any theory containing a massless spin 2 boson and a particle $X$ of spin $j$ must include a coupling
$A_3\left(1^{+2},2_X^{+j},3_X^{-j}\right)$ with the universal coupling
constant. Taking the soft limit of the helicity $+j$ particle, 
this coupling  has $\mathtt{a} = 2j-1$. As discussed in Section III, the condition for the vanishing of poles
in $z$ has no non-trivial solutions for $\mathtt{a}>3$, 
which implies $j\leq 2$. 
Thus, by demanding locality and unitarity in the soft limit,  
we find that massless higher spin particles cannot 
interact in any way with particles of spin $\leq 2$
in a theory of gravity. 
This is an on-shell version of the gravitational
Weinberg-Witten theorem \cite{Weinberg:1980kq}.

\vspace{1mm}
\noindent {\bf Supergravity.}  
We learned above that the usual soft graviton theorems arise from interactions with $\mathtt{a}=3$, for which $h_P = - h_k$. 
Such interactions include the standard graviton self interactions, 
and if we have spin 3/2 massless fields, the equivalence principle implies that the coupling of 
$A_3(1^{+2}, 2^{+\frac{3}{2}}, 3^{-\frac{3}{2}})$ 
must be the same as that of 
$A_3(1^{+2}, 2^{+2}, 3^{-2})$. Let us explore the soft 
limit of a positive-helicity spin 3/2 particle. With $h_s =\tfrac{3}{2}$, 
the interaction $A_3(1^{+2}, 2^{+\frac{3}{2}}, 3^{-\frac{3}{2}})$ has  $\mathtt{a}=2$, for which \reef{Ann} yields a non-trivial constraint from the absence of a $1/z$ pole. Consider for example 
$
  A_{n+1} \big(s^{+\frac{3}{2}}, 1^{-\frac{3}{2}}, 2^{+2}, 3^{+2},  4^{-2}, \ldots , n^{-2}\big) \,.
$
The $1/z$-pole in \reef{Ann} has three contributions with $\mathtt{a}=2$, namely from $k=1,2,3$.  (Lines $k=4,\ldots,n$ give  $\mathtt{a}=6$.)   
 The sum over the three channels $k=1,2,3$ gives the consistency condition
\be
  \begin{split} \label{swi}
  0=&\,[s1] A_n\big(1^{-2}, 2^{+2}, 3^{+2},  4^{-2}, \ldots, n^{-2}\big)\\
  &- [s2] A_n\big(1^{-\frac{3}{2}}, 2^{+\frac{3}{2}}, 3^{+2} , 4^{-2}, \ldots, n^{-2}\big)\\
  &- [s3] A_n\big(1^{-\frac{3}{2}}, 2^{+2}, 3^{+\frac{3}{2}},  4^{-2}, \ldots, n^{-2}\big)\,.
  \end{split}
\ee
This is precisely the MHV version of the $N=1$ supersymmetric Ward identities \cite{Grisaru:1976vm,Grisaru:1977px} (see also \footnote{The supersymmetric Ward identities have previously been derived from soft limits \cite{Grisaru:1977kk}, though assuming supersymmetry.}). Thus we reached the well known conclusion that spin 3/2 to couple to gravity supersymmetrically. The role of the usual reference spinor in the SUSY Ward identities is here played by the soft momentum.

\begin{acknowledgments}
We thank Thomas Dumitrescu for discussions which initiated this project. 
We also thank Ratin Akhoury for useful discussions. 
HE was supported in part by the US Department of Energy under Grant No. DE-SC0007859.
CRTJ was supported in part by an award to HE from the LSA Associate Professor Support Fund at the University of Michigan. SGN is supported by the National Science Foundation under Grant No.~PHY14-16123, and he also acknowledges sabbatical support from the Simons Foundation (Grant No.~342554 to Stephen Naculich). 
SGN also thanks the Michigan Center for Theoretical Physics
and the Physics Department of the University of Michigan
for generous hospitality
and for providing a welcoming and stimulating sabbatical environment.
\end{acknowledgments}

\appendix

\bibliography{softbib}% Produces the bibliography via BibTeX.

\end{document}